\documentclass[twocolumn]{aastex62}

\newcommand{\ms}{{m s$^{-1}$}}

\newcommand{\geff}{$g_{\mathrm{eff}}$}
\newcommand{\ewk}{EW$_{K}$}
\newcommand{\fwhmk}{FWHM$_{K}$}
\newcommand{\ma}{m\AA}
\usepackage[utf8]{inputenc} 
\usepackage[T1]{fontenc} 

\graphicspath{{./}{figures/}}

\revised{\today}

\newcommand{\PSUAA}{Department of Astronomy \& Astrophysics, 525 Davey Laboratory, The Pennsylvania State University, University Park, PA, 16802, USA}
\newcommand{\PSUCEHW}{Center for Exoplanets and Habitable Worlds, 525 Davey Laboratory, The Pennsylvania State University, University Park, PA, 16802, USA}

\newcommand{\UA}{Steward Observatory, The University of Arizona, 933 N.\ Cherry Ave, Tucson, AZ 85721, USA}

\newcommand{\STScI}{Space Telescope Science Institute, 3700 San Martin Dr, Baltimore, MD 21218, USA}
\newcommand{\JHU}{Department of Physics and Astronomy, Johns Hopkins University, 3400 N Charles St, Baltimore, MD 21218, USA}

\newcommand{\Macquarie}{Department of Physics and Astronomy, Macquarie University, Balaclava Road, North Ryde, NSW 2109, Australia }
\newcommand{\NIST}{National Institute of Standards \& Technology, 325 Broadway, Boulder, CO 80305, USA}
\newcommand{\CUBoulder}{Department of Physics, 390 UCB, University of Colorado, Boulder, CO 80309, USA}
\newcommand{\JPL}{Jet Propulsion Laboratory, California Institute of Technology, 4800 Oak Grove Drive, Pasadena, California 91109}

\newcommand{\UCI}{Department of Physics \& Astronomy, The University of California, Irvine, Irvine, CA 92697, USA}
\newcommand{\Carleton}{Carleton College, One North College St., Northfield, MN 55057, USA}

\newcommand{\Princeton}{Department of Astrophysical Sciences, Princeton University, 4 Ivy Lane, Princeton, NJ 08540, USA}
\newcommand{\RUSSELL}{Henry Norris Russell Fellow}

\newcommand{\MCDonald}{McDonald Observatory and Department of Astronomy, The University of Texas at Austin, 2515 Speedway, Austin, TX 78712, USA}
\newcommand{\UTSpace}{Center for Planetary Systems Habitability, The University of Texas at Austin, 2515 Speedway, Austin, TX 78712, USA}
\newcommand{\ONMCTIC}{Observat\'{o}rio Nacional/MCTIC, {R.} {Gen.} Jos\'{e} Cristino, 77, 20921-400 Rio de Janeiro, Brazil}
\newcommand{\AFRL}{Space Vehicles Directorate, Air Force Research Laboratory, 3550 Aberdeen Ave. SE, Kirtland AFB, NM 87117, USA}
\newcommand{\NOIRLab}{NSF's NOIRLab, 950 North Cherry Avenue, Tucson, AZ 85719, USA}
\shorttitle{Rotational Modulation of Zeeman Signatures}
\shortauthors{Terrien et al.}

\begin{document}

\title{Rotational modulation of spectroscopic Zeeman signatures in low-mass stars}

\correspondingauthor{Ryan Terrien}
\email{rterrien@carleton.edu}

\author[0000-0002-4788-8858]{Ryan C Terrien}
\affil{\Carleton}

\author[0000-0002-9800-9868]{Allison Keen}
\affil{\Carleton}

\author[0000-0001-9770-7624]{Katy Oda}
\affil{\Carleton}

\author[0000-0001-7142-2997]{Winter Parts}
\affil{\PSUAA}
\affil{\PSUCEHW}

\author[0000-0001-7409-5688]{Guðmundur Stefánsson}
\affil{\Princeton}
\affil{\RUSSELL}

\author[0000-0001-9596-7983]{Suvrath Mahadevan}
\affil{\PSUAA}
\affil{\PSUCEHW}

\author[0000-0003-0149-9678]{Paul Robertson}
\affil{\UCI}

\author[0000-0001-8720-5612]{Joe P.\ Ninan}
\affil{\PSUAA}
\affil{\PSUCEHW}

\author[0000-0001-7708-2364]{Corey Beard}
\affil{\UCI}

\author[0000-0003-4384-7220]{Chad F.\ Bender}
\affil{\UA}

\author[0000-0001-9662-3496]{William D. Cochran}
\affil{\MCDonald}
\affil{\UTSpace}

\author[0000-0001-6476-0576]{Katia Cunha}
\affil{\UA}
\affil{\ONMCTIC}

\author[0000-0002-2144-0764]{Scott A.\ Diddams}
\affil{\NIST}
\affil{\CUBoulder}

\author[0000-0002-0560-1433]{Connor Fredrick}
\affil{\NIST}
\affil{\CUBoulder}

\author[0000-0003-1312-9391]{Samuel Halverson}
\affil{\JPL}

\author[0000-0002-1664-3102]{Fred Hearty}
\affil{\PSUAA}
\affil{\PSUCEHW}

\author[0000-0001-5064-9973]{Adam Ickler}
\affil{\Carleton}

\author[0000-0001-8401-4300]{Shubham Kanodia}
\affil{\PSUAA}
\affil{\PSUCEHW}

\author[0000-0002-2990-7613]{Jessica E. Libby-Roberts}
\affil{\PSUAA}
\affil{\PSUCEHW}

\author[0000-0001-8342-7736]{Jack Lubin}
\affil{\UCI}

\author[0000-0001-5000-1018]{Andrew J. Metcalf}
\affil{\AFRL}
\affil{\NIST}
\affil{\CUBoulder}

\author[0000-0002-3628-4426]{Freja Olsen}
\affil{\Carleton}

\author[0000-0002-4289-7958]{Lawrence W. Ramsey}
\affil{\PSUAA}
\affil{\PSUCEHW}

\author[0000-0001-8127-5775]{Arpita Roy}
\affil{\STScI}
\affil{\JHU}

\author[0000-0002-4046-987X]{Christian Schwab}
\affil{\Macquarie}

\author[0000-0002-0134-2024]{Verne V. Smith}
\affil{\NOIRLab}

\author[0000-0001-5211-9458]{Ben Turner}
\affil{\Carleton}



\begin{abstract}
Accurate tracers of the stellar magnetic field and rotation are cornerstones for the study of M dwarfs and for reliable detection and characterization of their exoplanetary companions. Such measurements are particularly challenging for old, slowly rotating, fully convective M dwarfs. To explore the use of new activity and rotation tracers, we examined multi-year near-infrared spectroscopic monitoring of two such stars---GJ 699 (Barnard's Star) and Teegarden's Star---carried out with Habitable Zone Planet Finder spectrograph. We detected periodic variations in absorption line widths across the stellar spectrum with higher amplitudes towards longer wavelengths. We also detected similar variations in the strength and width of the 12435.67 {\AA} neutral potassium (K I) line, a known tracer of the photospheric magnetic field. Attributing these variations to rotational modulation, we confirm the known $145\pm15$~d rotation period of GJ 699, and measure the rotation period of Teegarden's Star to be $99.6\pm1.4$~d. Based on simulations of the K~I line and the wavelength-dependence of the line width signal, we argue that the observed signals are consistent with varying photospheric magnetic fields and the associated Zeeman effect. These results highlight the value of  detailed line profile measurements in the near-infrared for diagnosing stellar magnetic field variability. Such measurements may be pivotal for disentangling activity and exoplanet-related signals in spectroscopic monitoring of old, low-mass stars.

\end{abstract}

\keywords{}


\section{Introduction} \label{sec:intro}
M dwarf stars are common throughout the Galaxy \citep{Henry2006}, but there are important open questions about their magnetic activity and exoplanetary systems. Measurements and modeling of magnetic activity on these stars are rapidly improving \citep{2021A&ARv..29....1K}, facilitating new insights into their magnetic dynamos, particularly across the transition from partially to fully convective structure \citep{1997A&A...327.1039C}. M dwarfs also commonly host exoplanets \citep{2015ApJ...807...45D, 2020MNRAS.498.2249H,Sabotta2021}. However, magnetic activity can confuse or mask low-level exoplanetary signals \citep[e.g.,][]{2014MNRAS.439.3094B,Lubin2021}, hindering the construction of a census that might reveal the fingerprints of different exoplanet formation or migration mechanisms \citep[e.g., the prevalence of Earth-sized planets around stars of different masses,][]{Burn2021}. Further, the stellar magnetic field plays an important role in star-planet interactions \citep{Zarka2007,Turnpenney2018,Vedantham2020,Mahadevan2021} and on planetary habitability \citep[e.g.,][]{Shields2016}.

Central observables for these areas of inquiry are the stellar rotation period \citep[e.g.,][]{Reiners2014,Newton2017,2020AJ....159...52M} and the various manifestations of the stellar magnetic field \citep{2021A&ARv..29....1K}. Stellar rotation can modulate the appearance of regions with differing magnetic fields. As these regions rotate into and out of view, they can drive strong signals in the stellar photometry \citep{Irwin2011} and spectroscopy \citep{2020ApJ...897..125R,Lafarga2021}. Surface magnetic fields can also be observed more directly through the Zeeman effect on magnetically sensitive spectral lines \citep{2021A&ARv..29....1K}. This approach has shown promise for M dwarfs, chiefly with the use of spectropolarimetry to target the strong polarization signature of Zeeman splitting  \citep[e.g.,][]{Morin2008}. 

The weaker signals of Zeeman splitting on observed line profiles in the intensity (i.e.,~Stokes $I$, unpolarized) spectra include line broadening and strengthening (intensification), and depend on the strength of the line, the magnetic sensitivity of the transition, and the Zeeman pattern \citep{Stift2003}. These effects have been leveraged to estimate the total magnetic fields of several M dwarfs \citep{2019A&A...626A..86S,Moutou2017}, and have also revealed apparent saturation behavior of the dynamo \citep{Reiners2009,2020AJ....159...52M}. Measurements of unpolarized Zeeman signatures have even revealed rotational modulation in a small number of cases \citep{Kochukhov2017,Klein2021}, and may contribute to the periodic variability observed in numerous lines in M dwarf spectra \citep{Lafarga2021}.

We describe here a detailed study on the morphology of rotationally modulated changes in the time-series spectra of two fully convective, slowly rotating M dwarfs: GJ 699 (Barnard's Star) and Teegarden's Star. We show that these variations are consistent with manifestations of Zeeman broadening and intensification under a changing magnetic field. These results provide the most detailed evidence yet that long-term monitoring of Zeeman broadening and intensification in the intensity spectra can provide a powerful diagnostic of the magnetic fields of fully-convective stars. Further, these results suggest a promising path for reliably identifying activity-induced variations embedded in precise radial velocity (RV) monitoring of exoplanet host targets.

\section{Methods} \label{sec:methods}
\subsection{Observations with HPF} \label{sec:observations}
We examined the time-series Habitable Zone Planet Finder (HPF) spectra of GJ 699 and Teegarden’s Star. HPF \citep{mahadevan2012,mahadevan2014} is a high-resolution ($R\sim55,000$), stabilized \citep{stefansson2016} spectrograph covering 820-1280~nm. Observing from the 10m Hobby-Eberly Telescope (HET) at the McDonald Observatory in Texas, HPF is optimized for RV surveys of northern mid-to-late M dwarf stars. HPF is wavelength-calibrated using a custom-built laser frequency comb (LFC) calibration system \citep{2019Optic...6..233M} and has a stable instrumental response function \citep{Kanodia2021}, enabling an RV precision of $\sim$1~{\ms} on-sky \citep[e.g.,][]{2021arXiv210507005L}.

GJ 699 (M4) and Teegarden's Star (M7) are routinely-observed standard stars for HPF, and thus have long and well-sampled baselines. Our sample of GJ 699 spectra consists of 1095 spectra (corresponding to 126 visits) exceeding a signal-to-noise ratio (SNR) of 100 (at 10,700~\AA, per reduced spectral pixel) and spanning 2019 April to 2021 July. Our sample of Teegarden’s Star spectra contains 155 spectra (82 visits) exceeding the same SNR threshold and spanning 2018 October to 2021 March. These spectra were produced by the standard HPF pipeline \texttt{HxRGproc}, which is described further in \cite{2018SPIE10709E..2UN}, \cite{2019ASPC..523..567K}, and \cite{2019Optic...6..233M}. We use a synthetic telluric spectrum \citep{2014AJ....148...53G} to mask out regions that are affected by telluric contamination. 

We evaluated the stability of individual absorption line properties (Section \ref{sec:features}), as well as average line widths (Section \ref{sec:dlw}) in these time-series spectra. 

\subsection{Line Properties} \label{sec:features}
\begin{figure}
    \includegraphics[width=0.51\textwidth]{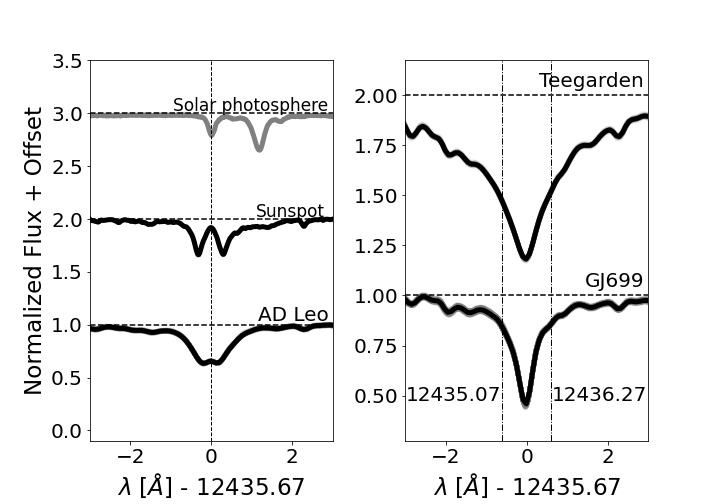}
    \caption{The K I line in multiple stellar spectra, with dashed lines representing the estimated continuum level. The left panel shows comparisons between the Solar photosphere \citep{Livingston1991,Wallace1993} and sunspot  spectra \citep{Wallace1992,Wallace1999}, and the highly-active M dwarf AD Leo (with $\sim$kG magnetic fields), and shows the clear presence of Zeeman splitting for this line. The right panel shows all spectra for each of our target stars overlaid in the vicinity of the K~I line, as well as the boundaries used for the EW calculation.}
    \label{fig:features}
\end{figure}

We examined the well-studied \citep{Fuhrmeister2021} 12435.67 {\AA} neutral potassium (K I) absorption line, shown in Figure \ref{fig:features}. We compiled line properties from the Vienna Atomic Line Database \citep[VALD,][]{Ryabchikova2015} (air wavelength: 12432.27 {\AA}, Transition: 4p\,$^2$P$^{\mathrm{o}}_{1/2}-$5s\,$^2$S$_{1/2}$, effective Land\'{e} factor {\geff} = 1.3, $\gamma_{\mathrm{radiative}}=7.74836 \times 10^8~\mathrm{rad}~\mathrm{s}^{-1}$, $\gamma_{\mathrm{Stark}}=1.65657 \times 10^8~\mathrm{rad}~\mathrm{s}^{-1}$ per $10^{12}$ perturbers per cm$^{3}$, and $\gamma_{\mathrm{vdW}}=43.57224 \times 10^8~\mathrm{rad}~\mathrm{s}^{-1}$ per $10^{16}$ perturbers per cm$^{3}$). This line is exceptionally strong and free of both significant telluric contamination and blending with strong neighboring lines. It exhibits the expected Zeeman splitting in the presence of a magnetic field, as shown in Figure \ref{fig:features} by comparison to Solar photospheric \citep{Livingston1991,Wallace1993} and sunspot \citep{Wallace1992,Wallace1999} spectra and in HPF spectra of the AD Leo (Figure \ref{fig:features}), an M dwarf with well-measured $\sim$kG magnetic fields \citep{2018MNRAS.479.4836L}. This behavior is further confirmed in CARMENES spectra of multiple stars by \citet{Fuhrmeister2021}, who also establish that this feature is insensitive to chromospheric variability and therefore a powerful indicator of photospheric properties, including magnetism.

This K~I feature appears in the second reddest HPF spectral order (order index 26, spanning approximately 12367--12525~\AA), which is largely clear of telluric contamination. To track the width of the K~I line, we measured {\fwhmk} directly, using the \texttt{scipy.signal.peak\_widths} function \citep{2020NatMe..17..261V}. To track the strength of this line, we measured the (pseudo-)equivalent width (\ewk) defined in the usual way:
\begin{equation}
    \mathrm{EW} = \sum_{\lambda_1}^{\lambda_2} ( 1 - F_{\lambda} / F_{c} ) d\lambda,
\end{equation}
where $\lambda_1$ and $\lambda_2$ are the wavelength limits of the feature (see Figure \ref{fig:features}), $F_\lambda$ is the stellar flux at wavelength $\lambda$, and $F_c$ is the continuum flux at $\lambda$. We account for fractional pixels at the feature boundaries using a linear interpolation across pixels, and our continuum flux level is defined by the HPF pipeline. In the spectra of GJ~699 and Teegarden's Star, the continuum level is reached at approximately $\pm2$~{\AA} and $\pm5$~{\AA}, respectively.

The K~I line is notably broader in the spectra of Teegarden's Star than in the warmer GJ 699; this results from the increasing depth of line formation with decreasing temperature, which reveals the increasing effect of collisional broadening with the H$_2$ molecule \citep[][and references therein]{Fuhrmeister2021}. Our simulations indicated that the impact of Zeeman intensification would at either temperature be localized to a region near the line core, so we elect to use uniform narrow boundaries for the {\ewk} calculation. This helps to isolate potential Zeeman-driven variability in the K~I line from other nearby spectral variations; tests with broader windows for Teegarden's star were nonetheless consistent with the results presented below.

We estimated the precision of our {\fwhmk} and {\ewk} measurements using a Monte Carlo approach and the HPF pipeline-reported variances for each 1-D pixel. By measuring each parameter on 100 different realizations of several different HPF spectra of GJ 699 and Teegarden's Star, we estimated the precision of our FWHM measurement at 10~{m\AA} ($\sim1$\%), and 2~{m\AA} for the EW measurement ($\sim0.5$\%). 

\subsection{Differential Line Width}\label{sec:dlw}
To assess the presence of Zeeman broadening, we also examined the average widths of the absorption lines in the stellar spectra using the differential line width indicator (dLW) as defined and implemented in the SpEctrum Radial Velocity AnaLyzer (SERVAL) pipeline \citep{2018A&A...609A..12Z}. To measure the dLWs for HPF spectra, we used the SERVAL pipeline optimized for HPF spectra  \citep{2019Optic...6..233M,2020AJ....159..100S}. Briefly, the SERVAL algorithm involves the construction of a high-SNR template and matching of individual observations against this template. The dLW results from considering the second (spectral) derivative of the template during the matching process, and encodes the combined line width changes across all lines used in the template. The dLW is analogous to the FWHM of the CCF, but presents a simpler profile in the context of M dwarf spectra which have many blended lines. In the calculation of the dLW, we mask out spectral regions impacted by tellurics and sky emission lines as discussed in \citep{2020AJ....159..100S}. We estimated the dLW uncertainty as described in \citet{2018A&A...609A..12Z}, and found a typical per-spectrum precision of 5~m$^{2}$s$^{-2}$ (5-10\%).

To study the dLW behavior as a function of wavelength, we also calculated the dLW for individual spectral orders.  Estimating the dLW per order enabled 10 independent measurements (corresponding to HPF's most telluric-free orders) of dLW spanning 840nm--1250nm.

\subsection{Independent Measurements of Differential Line Width}
Teegarden's star has also been monitored extensively with CARMENES \citep{Zechmeister2019}, which has visible ($R\sim$94600, 5200-9600\AA) and NIR ($R\sim80400$, 9600-17100\AA) channels. Public data used here consist of 244 (in the NIR channel, 245 in the visible channel) measurements of the dLW, spanning an observing baseline from 2016 to 2019 and with typical SNR of 58 per extracted pixel around 746~nm. 

\subsection{Periodicity Measurement}
To measure the periodicity of modulations present in our spectroscopic indicators, we employed the generalized Lomb-Scargle (GLS) periodogram \citep{Zechmeister2009}, as implemented in \texttt{astropy} \citep{2013A&A...558A..33A,2018AJ....156..123A}. We searched frequencies corresponding to periods of 3-500~d, using a frequency grid sufficiently dense to provide 10 samples across a given periodogram peak. We estimated the false alarm probability using the method described in \citet{Baluev2008}, following similar recent work \citep{Lafarga2021}.\footnote{We note that the interpretation of periodogram peak significance is not necessarily straightforward, and we refer the reader to \citet{VanderPlas2018} and references therein for a discussion of the related caveats and implications.}

\subsection{Simulations}
\label{section:simulations}
To assess whether the observed variations were consistent with variable Zeeman splitting, we simulated the K~I line using the \texttt{NICOLE} \citep{SocasNavarro2015} program, with plane-parallel MARCS atmosphere \citep{Gustafsson2008} inputs corresponding approximately to GJ 699 and Teegarden's Star (3300~K, [Fe/H]=0.0 and 2900~K, [Fe/H]=-0.25 respectively, and $\log(g)=5.0$, no microturbulence) and line parameters extracted from VALD (given above). 

For each star, we manually adjusted the K abundance to approximately match the depth of the observed line profile. For Teegarden's Star, we used [K]=4.92\footnote{Abundances quoted on the usual logarithmic scale as $\log(N_{K}/N_H) + 12.0$.}, and for GJ 699, we used [K]=4.62. These values are generally consistent with the old age and sub-solar metallicity estimates for both stars \citep{Muirhead2012,Zechmeister2019}, though we emphasize that our goal in this work is not precise abundance determination, but to look at temporal variability of the line. We set the macroturbulent velocity to 1~km~s$^{-1}$ for both stars, in order to approximately match the observed line widths.  For simplicity, we assumed local thermodynamic equilibrium (LTE) in these simulations. We caution that departures from LTE are expected for other K~I lines in cool stars \citep[e.g.,][]{Reggiani2019} and may impact the comparison between simulated and observed spectra, although \citet{Reggiani2019} found minimal departures from LTE for the low temperatures and high log(g) values associated with M dwarfs. 

We implemented a radially-oriented two-component magnetic field strength distribution, described by a magnetic field strength $B$ and a filling fraction $f$ \citep{2021A&ARv..29....1K}, which are uniform for all simulated lines of sight. 

We disk-integrated over the visible hemisphere of the star by first simulating in \texttt{NICOLE} a series of 512 1-dimensional spectra along a line from disk center to disk edge. These spectra were evenly spaced in projected radius from disk center, and account for the effect of limb darkening as well as the changing angle between the magnetic field and the line of sight. To account for stellar rotational broadening, we applied appropriate Doppler shifts to each line of sight assuming rigid-body rotation perpendicular to the line-of-sight and rotational periods corresponding to the measured period for each star (see Section \ref{sec:rotationperiod}). We then formed the disk-integrated line profile by integrating over the rotating and limb-darkened lines of sight. 

\section{Results}

We detected significant periodic variations in the K~I widths and strengths ({\fwhmk}, {\ewk}) and in the general line widths (dLWs), consistent with rotational modulation of Zeeman intensification and Zeeman broadening. We detail these results, and their interpretation, below.

\subsection{Spectroscopic Measurements}

\begin{figure*}
    \includegraphics[width=1\textwidth]{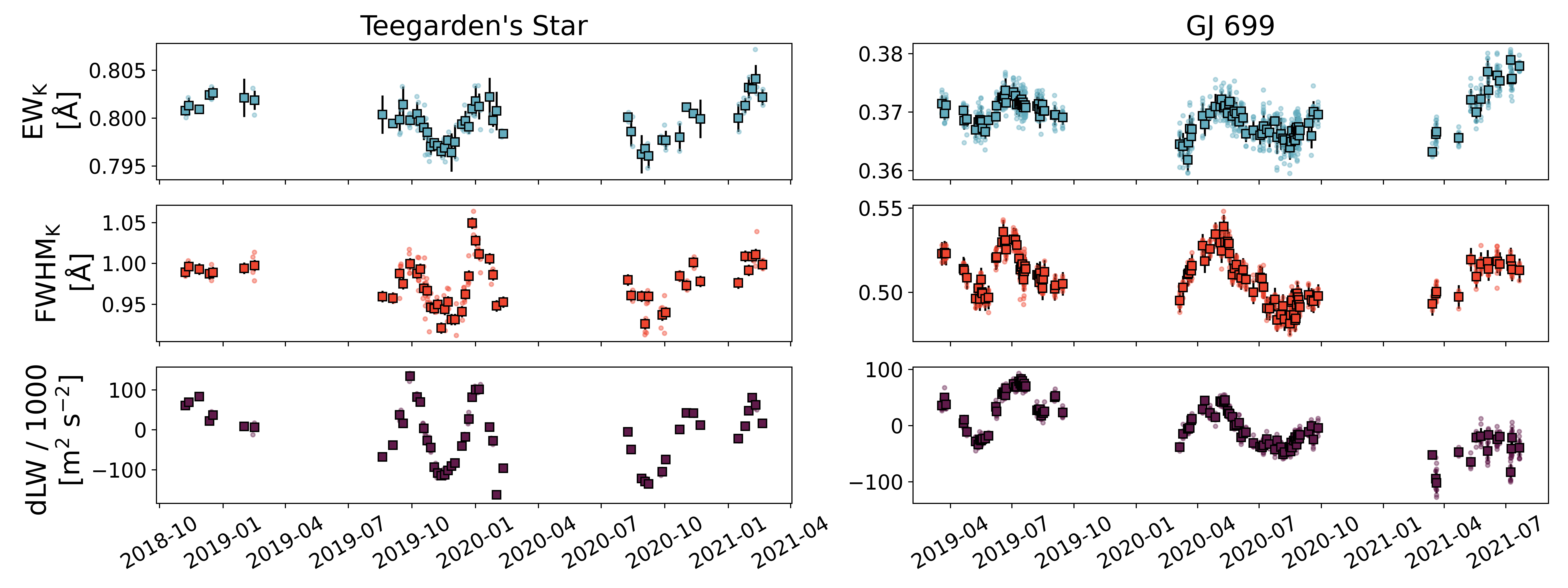}
    \caption{\ewk (top), \fwhmk (middle), and dLW (bottom) as measured for Teegarden's star (left) and GJ 699 (right) with HPF. Individual measurements are shown as points, while boxes show binned averages (daily for GJ~699, five days for the more sparsely sampled Teegarden's star) with error bars corresponding to either 1) the measured scatter within that day if multiple spectra are obtained, or 2) the baseline single measurement precision estimate if only one spectrum was taken during the relevant bin. Significant periodic variations are present in all measurements, and are largely in-phase across these separate measurements.}
    \label{fig:timeseries}
\end{figure*}

The time series measurements of {\fwhmk}, {\ewk}, and dLW in GJ 699 and Teegarden's star are shown in Figure \ref{fig:timeseries}, and reveal significant, similar, long-term variations over $>2$~yr. {\fwhmk} varies at the 50-100~{\ma} level ($\sim10\%$), {\ewk} varies at the 5-10~{\ma} level ($0.5$-$3$\%), while the dLWs vary at $\gtrsim100$~m$^{2}$s$^{-2}$. The amplitudes of these variations exceed the estimated per-spectrum uncertainties.

\begin{figure*}
    \includegraphics[width=1\textwidth]{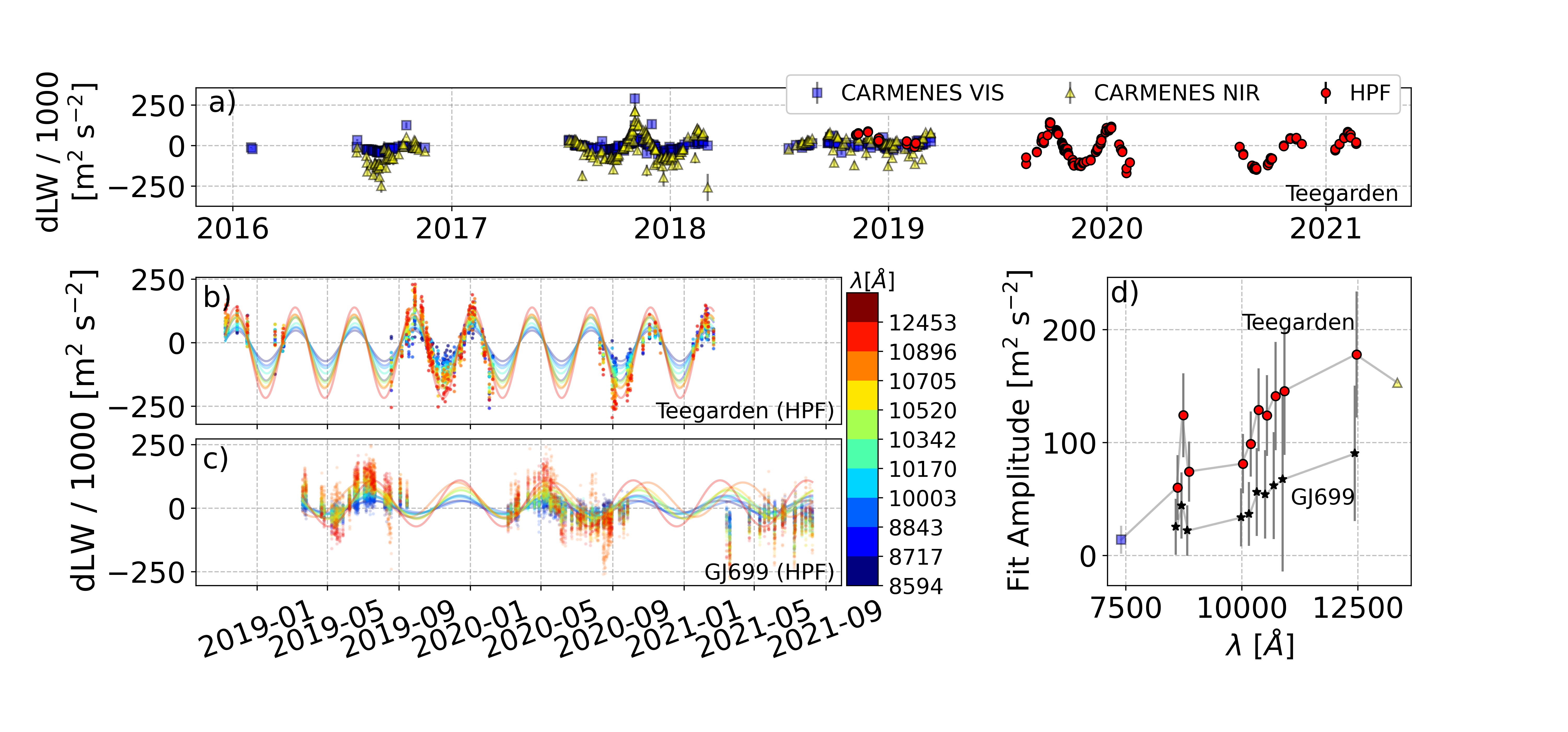}
    \caption{a) Time series dLW measurements for Teegarden's Star from HPF and CARMENES visible (VIS) and infrared (NIR) channels \citep{Zechmeister2019}. Periodic modulations are detected by both instruments, and are larger in amplitude at infrared wavelengths (HPF, CARMENES NIR) than at visible wavelengths (CARMENES VIS). b) The HPF time series dLW measurements for Teegarden's Star, separated and color-coded by spectrograph order. Redder (bluer) wavelengths are longer (shorter). Overplotted are the best-fit sinusoids for each wavelength region as determined by the GLS package \citep{Zechmeister2009}. c) The same as b), but for GJ 699. A consistent oscillatory pattern is clear at each wavelength, and longer wavelengths tend to correspond to higher amplitudes of variation. d) The amplitudes of the best-fit sinusoids to the order-by-order dLW measurements for Teegarden's star (circles) and GJ 699 (stars) as a function of wavelength, showing increasing amplitude with wavelength as expected for the Zeeman effect. The black circles for Teegarden's star are calculated from CARMENES data: the left point (CARMENES-VIS) is taken from a best-fit sinusoid; the amplitude from the noiser CARMENES-NIR data is simply estimated as half of the difference between the 2 and 98th percentiles. Error bars for all best-fit amplitudes represent the residual RMS after the removal of the best-fit sinuosid.}
    \label{fig:teegarden_dlw_chromatic}
\end{figure*}

Figure \ref{fig:teegarden_dlw_chromatic} shows the dLW measurements in detail. CARMENES data dramatically extends the dLW baseline for Teegarden's star. CARMENES data from both the visible and NIR channels shows a long-term periodicity, which is consistent with the variations detected with HPF. This is discussed in more detail in Section \ref{sec:discussion}.

We also examined the dLW measurements at different wavelengths by comparing the HPF and CARMENES dLW values (for Teegarden's star), as well as the per-order dLW measured in the HPF spectra (Figure \ref{fig:teegarden_dlw_chromatic}). A trend of increasing variability amplitude with increasing wavelength is apparent for both stars in the per-order dLW values, and between instruments for Teegarden's star. As discussed below, this increasing amplitude with wavelength is expected of Zeeman broadening \citep{Reiners2012}.

\begin{figure*}
\gridline{\hspace{-1cm}\fig{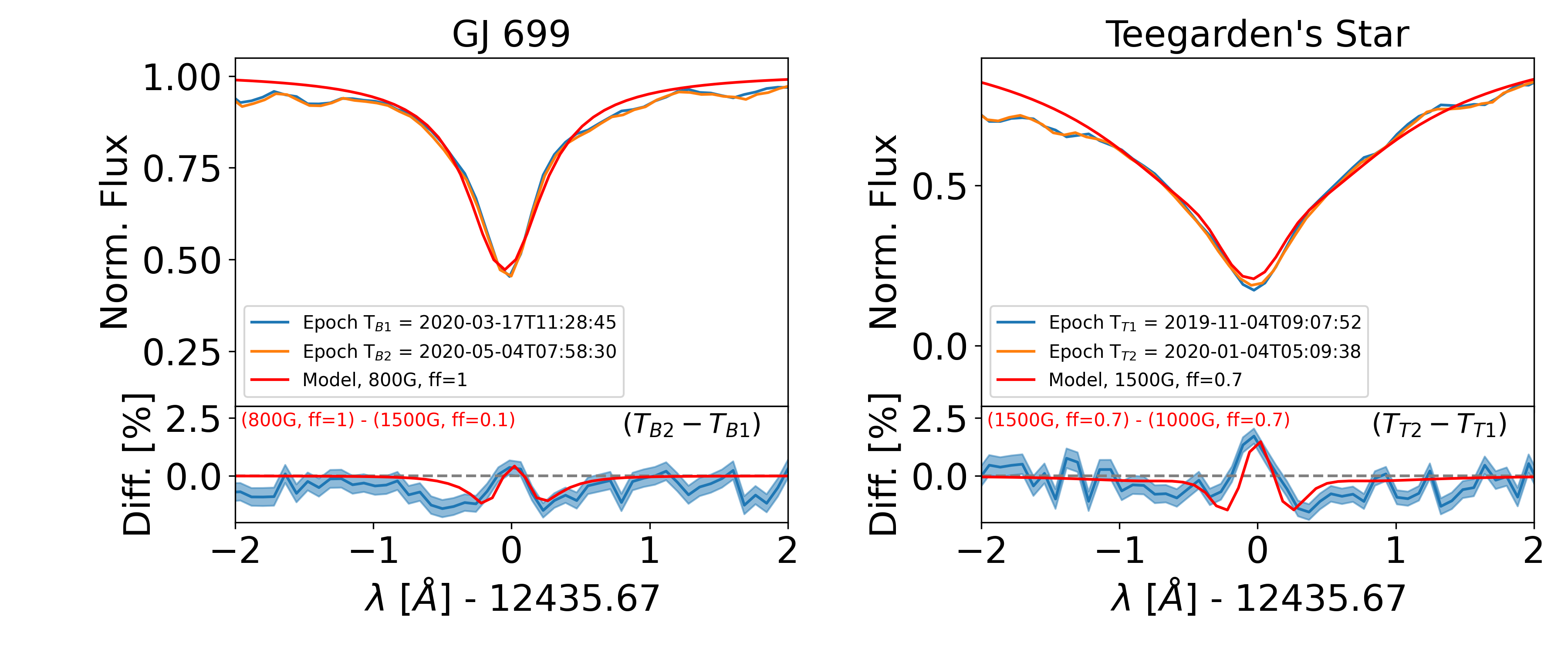}{.9\textwidth}{(a)}}
\gridline{\fig{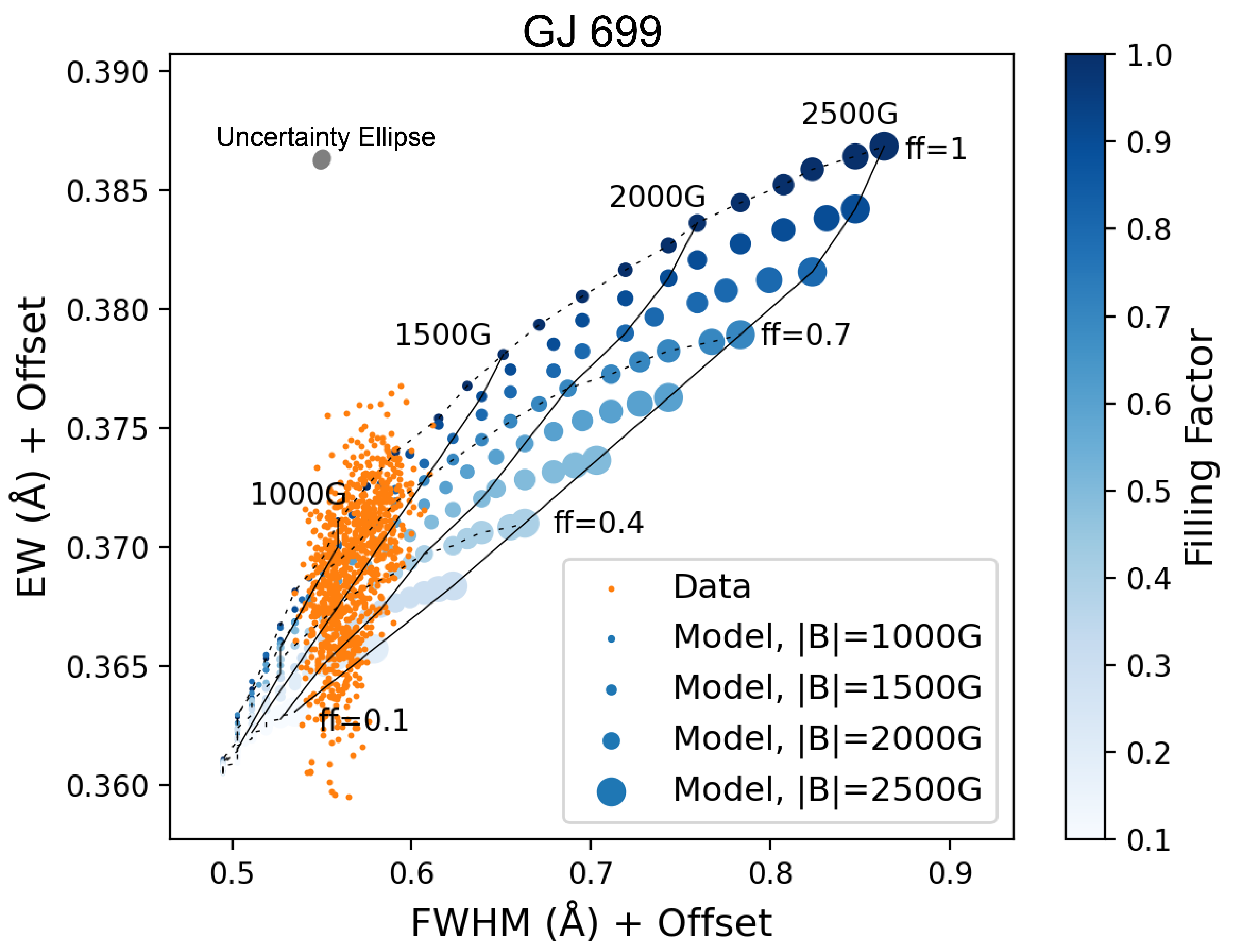}{0.45\textwidth}{(b)}
          \fig{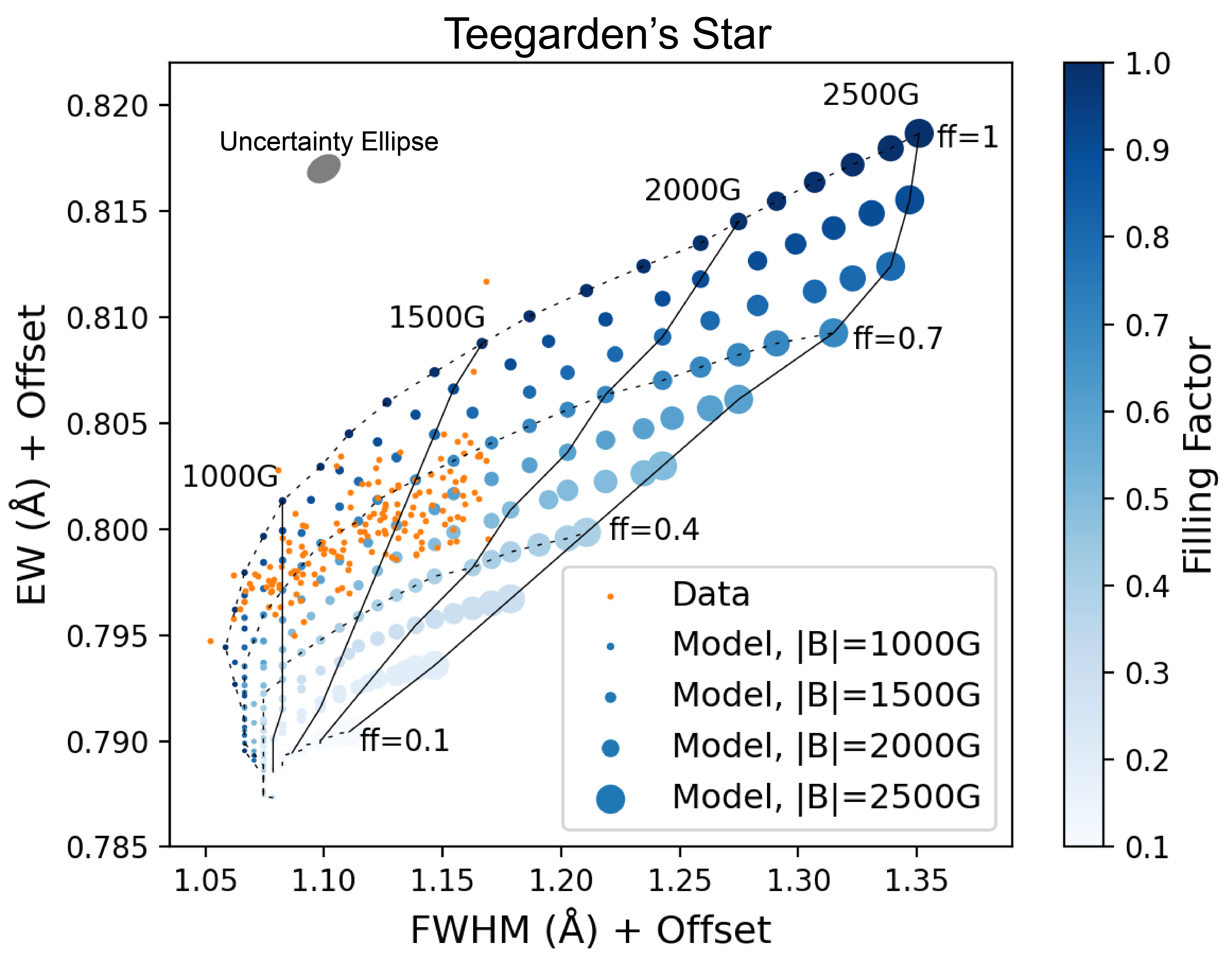}{0.45\textwidth}{(c)}
          }
\caption{(a) The detailed line profile of the K~I line in GJ 699 and Teegarden's Star, showing the change between epochs of low and high \ewk. In each case, epoch 1 (2) corresponds to a period of low (high) \ewk, \fwhmk, and dLW. The difference, shown in the bottom panels, reveals a line shape change (from first to second epochs) that is consistent with a line that is strengthening and broadening under the Zeeman effect (simulated difference overlaid). (b,c) Measured (orange) and simulated (blue) variations over time of K~I line {\ewk} and {\fwhmk} for (b) GJ 699 and (c) Teegarden's Star. Simulations were generated using \texttt{NICOLE} \citep{SocasNavarro2015} as described in Section \ref{section:simulations}, varying the magnetic field strength and filling factor. In both stars, the approximate range of variation observed for {\ewk} and {\fwhmk} can be reproduced in the simulated spectra. Gray ovals represent the approximate error ellipse, accounting for typical pipeline-reported variances in the measured HPF spectra for each star.
\label{fig:lineshape}}
\end{figure*}

To gain insight into the behavior of the K~I line, we also examined the line shape change between epochs of low and high {\ewk}, as shown in Figure \ref{fig:lineshape}. This figure confirms that higher {\ewk} correlates with a broader line profile (as suggested in Figure \ref{fig:timeseries}), and also with a shallower line core. 

We also explored whether the variation observed in {\ewk} and {\fwhmk} is consistent with what might be produced by rotational modulation of magnetic features on the surfaces of GJ 699 and Teegarden's Star, using the simulations described in Section~\ref{section:simulations}. Figure \ref{fig:lineshape} shows the resulting variation of the line shape parameters when the magnetic field strength and filling factor of our two-component model are varied. We note that the {\ewk} and {\fwhmk} are dependent on the precise choice of continuum level, which cannot be defined equivalently in the observed and simulated spectra, because the latter has only a single source of opacity. We therefore include an offset for each of these quantities to align the simulated and measured values, and emphasize that the relevant comparison here is the range of variation.

For both stars, we observe that the range of variation and the measured line shapes themselves can be approximately reproduced by magnetic field strength and filling factor variations in the simulated spectra (Figure \ref{fig:lineshape}). We note that our two-component model is likely a significant oversimplification of the true magnetic field configuration, and that there are numerous degeneracies between the line profile measurements and the line parameters adopted for these simulations which complicate the reliable determination of a single magnetic field and filling factor for a given spectrum.\footnote{One complication is between the absorbing species abundance and magnetic field strength; properly disentangling these two properties can be accomplished, for example, if another absorption line from the same species is accessible and has {\geff=0.} \citep[e.g.,][]{2019A&A...626A..86S}.} Nonetheless, we conclude that the range of variations observed in our line profile measurements can at least be plausibly explained by variations in our magnetic field model.

\subsection{Rotation Period Measurements}
\label{sec:rotationperiod}
\begin{figure}
    \includegraphics[width=0.53\textwidth]{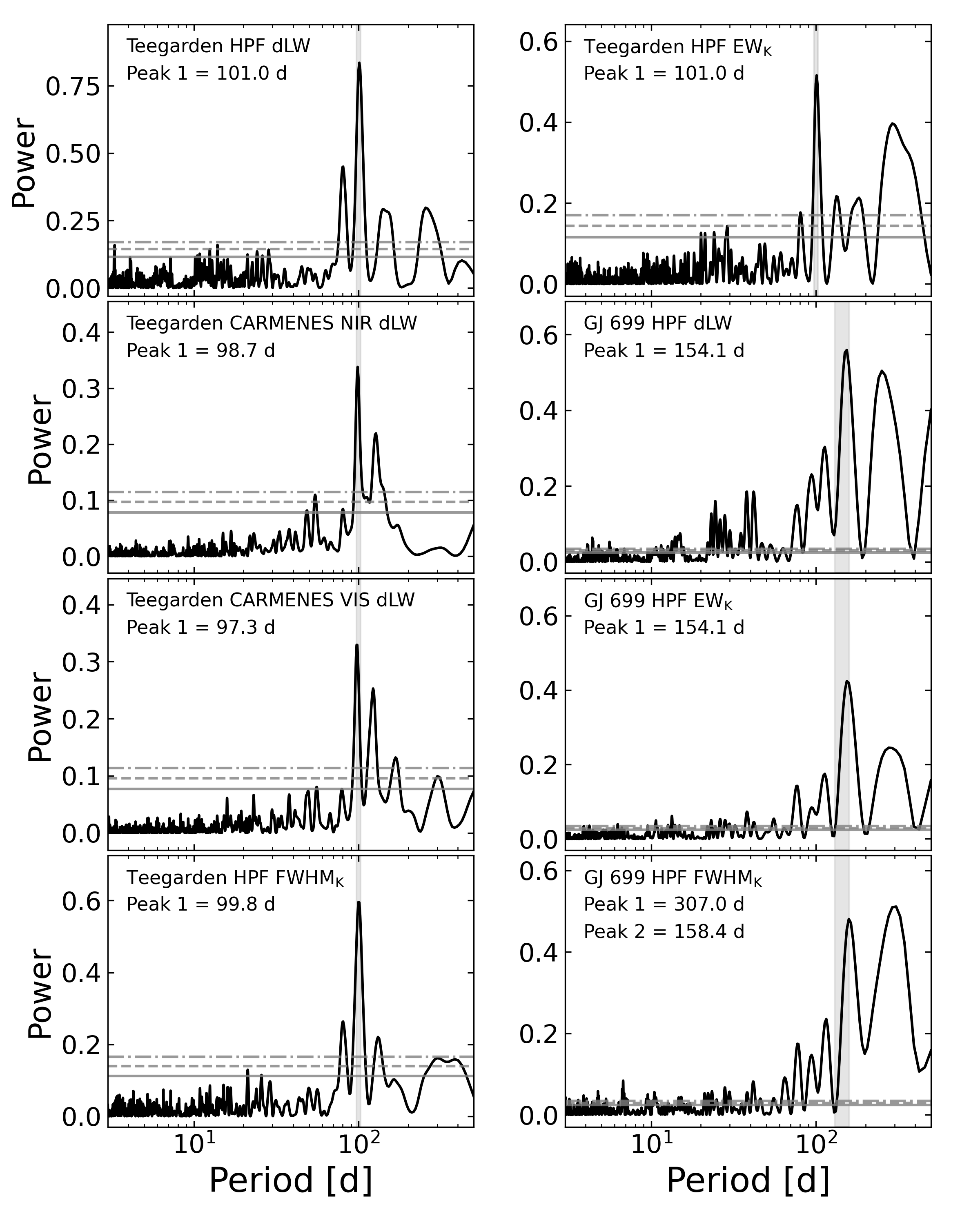}
    \caption{GLS Periodograms for Teegarden's Star and GJ 699. In each case, the False Alarm Probability estimates for 10\% (solid line), 1\% (dashed line), and 0.1\% (dash-dot line) are indicated. For Teegarden's star, a $99.6\pm1.5$~d periodicity (highlighted) is strong in each of the dLW measurements (HPF, and CARMENES VIS and NIR), as well as in the \ewk and \fwhmk measurements. For GJ 699, each indicator has a strong peak at the $145\pm15$~d rotation period \citep[highlighted,][]{2019MNRAS.488.5145T}.}
    \label{fig:periodograms}
\end{figure}

The GLS periodogram results for our measurements, including the periods of the most significant peaks, are shown in Figure \ref{fig:periodograms}. For GJ 699, the most recent observing season appears qualitatively different from the previous two; this change may be related to a recently-uncovered detector bias voltage drift. For completeness we retain these data points in our dataset, but for our period analysis we conservatively use only the data through 2020.

For Teegarden's star, the spectroscopic indicators are unanimous in the detection of a strong periodicity. We estimate the period by taking the mean and standard deviation of the periodogram peaks from the five time series (HPF dLW, Carmenes NIR and VIS dLW, HPF {\fwhmk}, and HPF {\ewk}), finding $99.6\pm1.4$~d, which we identify with the rotation period.

For GJ 699, we found strong peaks in the periodograms near the suspected $145\pm15$~d rotation period of this star \citep{2019MNRAS.488.5145T}. Among the HPF time series, the mean and standard deviation of these peak values are 155~d and 2~d, respectively. We note that the time baseline of our dataset is smaller than that of \citet{2019MNRAS.488.5145T}, who find a forest of periodogram peaks for GJ 699 at $145\pm15$~d which may be related to differential rotation. The periodicity detected in our dataset is more tightly localized around 155~d; this may be an artifact of the shorter time baseline. Another significant peak in our periodograms around 300~d may be a harmonic of the rotation period, or due to a longer-term activity cycle.

\section{Discussion}
\label{sec:discussion}

These measurements offer some of the clearest examples yet of rotationally-modulated variations in the spectra of slowly-rotating, fully-convective stars. These measurements enable the recovery of a signal near the rotation period of $145\pm15$~d of GJ 699 \citep{2019MNRAS.488.5145T}, and allow us to measure the rotation period of Teegarden's Star at $99.6\pm1.5$~d. Notably, dLW modulations at $\sim100$~d are clearly present in the data of \citet{Zechmeister2019}, who caution against over-interpreting these variations due to potential contamination from tellurics, instrumental focus changes, and other effects. The consistency of the signal between CARMENES and HPF suggests that it is indeed of stellar origin. Further, this relatively long rotation period is consistent with an old age indicated by the low H$\alpha$ emission of Teegarden's Star \citep{Zechmeister2019}.

\subsection{Magnetic Origin}
We argue that the observed variations are consistent with a changing magnetic field imprinting on the spectra via the Zeeman effect. The indicators ({\fwhmk}, {\ewk}, and dLW) change in phase with each other (Figure \ref{fig:timeseries}), and at levels consistent with our simple model of a changing surface magnetic field. Further, the line width changes are stronger at longer wavelengths (Figure \ref{fig:teegarden_dlw_chromatic}), as is expected for Zeeman broadening, in which the wavelength separation of the split line components ($\Delta \lambda$) from the unsplit line center ($\lambda_0$) behaves as
\begin{equation}
\Delta \lambda \propto g_{\mathrm{eff}} B \lambda_0^2,
\end{equation}
where {\geff} is the effective Land\'e factor and $B$ is the magnetic field \citep{2021A&ARv..29....1K}. Finally, the line shape change of the K~I line is consistent with simple simulations of the impact of a changing magnetic field (Figure \ref{fig:lineshape}).

We note, however, the potential contribution of other effects on the stellar surface. For example, stellar rotation that brings regions of varying magnetic field strength into and out of view may also modulate the appearance of associated bright/hot regions (faculae) or dark/cool regions (spots). Spot and faculae sizes and filling fractions are not well-constrained for these M dwarfs \citep[e.g.,][]{Rackham2018}, and we cannot rule out that the combined effects of these regions may be the origin of some of the rotational modulation we observe. However, we do note that in simple (single temperature) simulations, the K~I line becomes both deeper and broader with decreasing temperature, rather than the stronger/shallower behavior shown in Figure \ref{fig:lineshape}. This favors the Zeeman explanation over a simple temperature contrast.

A fully self-consistent description of the effect of magnetic fields on the stellar surface would account for the impact of the magnetic fields on the gas dynamics via magneto-hydrodynamic simulation. Indeed, the resulting changes of line positions and profile asymmetries are expected to be an important pathway for stellar activity to print through to RV measurements \citep{Meunier2017}, although the dynamics may be somewhat different for M dwarfs than for FGK stars \citep{Liebing2021}. With only the K~I line examined in detail, we cannot measure the low-level line asymmetries which accompany changes in the visible convective pattern, and so cannot constrain if or how this may impact our measurements. However, we do note that \citet{2019MNRAS.488.5145T} examined the bisector behavior in GJ 699 and found no detectable signal of changing line asymmetry at their level of measurement precision.

Finally, it is possible that the measured line strengths are also impacted by variable chromospheric emission, which is thought to drive some of the variability in line-by-line studies of correlation with established activity indicators \citep[e.g.,][]{Wise2018}. However, \citet{Fuhrmeister2021} consider correlations between the established chromospheric H$\alpha$ activity indicator and the K~I line for a range of M dwarfs and find that it exhibits no measurable chromospheric component (in contrast to K~I lines at shorter wavelengths). Further, pure chromospheric ``filling-in" of the line core would not be expected to exhibit the tight correlation observed between {\ewk} and {\fwhmk}, nor would we expect it to be so widespread across the spectrum as indicated by the dLW measurements.

\subsection{Implications}
The strength of the rotational modulations stands in contrast to the low photometric variability in both GJ 699 and Teegarden's Star. Careful analysis revealed no rotation period signal in the photometry of Teegarden's star by \citet{Zechmeister2019}, and the rotation period was only marginally detected in some subsets of photometric data used by \citet{2019MNRAS.488.5145T}. The clear spectroscopic detection of rotation in both stars echos the findings of \citet{Lafarga2021}, who show that the rotation period is frequently recoverable in the cross-correlation function parameters, chromospheric indicators, and dLW of M dwarfs. These findings indicate that spectroscopic monitoring may often provide the best handle on the rotation period for these types of stars. 

The need to disentangle stellar activity from exoplanetary signals in RV data is well-recognized, and has motivated careful statistical \citep[e.g.,][]{Davis2017} and line-by-line \citep[e.g.,][]{Dumusque2018,Wise2018} analyses to tease these effects apart. To the extent that our observed modulations are driven by the Zeeman effect, this provides a compelling justification for spectroscopic monitoring of RV targets (especially fully convective M dwarfs) at infrared wavelengths, where the Zeeman impact is largest. Together with the results of \citet{Lafarga2021}, the K~I line studied here appears to provide a powerful indicator of the changing photospheric magnetic field on these stars, which may be expected to perform well as a tracer of rotational modulation, and as an RV activity indicator.

Recently, \citet{Klein2021} used Zeeman Doppler Imaging to recover the magnetic field geometry of Proxima Centauri, an old, slowly-rotating, fully-convective star similar to GJ 699 and Teegarden's Star (albeit more active). They additionally found a significant rotational modulation in Stokes I spectra of Proxima Cen, and discuss the value of contrasting polarized and unpolarized indicators for diagnosing the properties of the dynamos of these fully-convective stars. Based on the significance of the unpolarized Zeeman signatures presented here, we suggest that long-term spectropolarimetric and spectroscopic monitoring of GJ 699 and Teegarden's Star may prove similarly revealing about the (differential) rotation, dynamo mechanisms, and activity cycles of fully convective stars.

\section{Conclusion}
We present the detection of significant rotational modulation of absorption line strength and width in the time-series spectra of the slowly-rotating, fully convective M dwarfs GJ 699 and Teegarden's Star. By establishing the wavelength-dependence of the line width signal, and by comparing to simulations of the expected line shape change, we argue that these signals are consistent with the expected effect of a changing magnetic field mediated by the Zeeman effect. Using the GLS periodogram technique to measure the periodicities in these signals, we also confirm the rotation period of GJ 699 at $145\pm15$~d, and establish the rotation period of Teegarden's Star at $99.6\pm1.4$~d. The use of these Zeeman tracers in high-resolution stabilized NIR spectra provides a promising approach for the extraction of rotation and magnetic field information for the detailed study of fully-convective stars and their exoplanatery companions. The promise of highly stabilized NIR spectrometers is now being realized in their ability to measure and potentially help mitigate even subtle stellar activity signatures.

\acknowledgments
We thank the anonymous reviewer for prompt feedback that improved the quality of this manuscript. 

These results are based on observations obtained with the Habitable-zone Planet Finder Spectrograph on the Hobby-Eberly Telescope. The Hobby-Eberly Telescope is a joint project of the University of Texas at Austin, the Pennsylvania State University, Ludwig-Maximilians-Universitat Munchen, and Georg-August Universitat Gottingen. The HET is named in honor of its principal benefactors, William P. Hobby and Robert E. Eberly. The HET collaboration acknowledges the support and resources from the Texas Advanced Computing Center. We thank the Resident astronomers and Telescope Operators at the HET for the skillful execution of our observations with HPF.

We acknowledge support from NSF grants AST 1006676, AST 1126413, AST 1310875, AST 1310885, AST 2009955, AST 2009889, AST 2009982, AST 2009554, and the NASA Astrobiology Institute (NNA09DA76A) in our pursuit of precision radial velocities in the NIR. We also acknowledge support from the Heising-Simons Foundation via grant 2017-0494. This research was conducted in part under NSF grants AST-2108493, AST-2108512, AST-2108569, and AST-2108801 in support of the HPF Guaranteed Time Observations survey. We also acknowledge support from the NASA Extreme Precision Radial Velocity program via grant 80NSSC21K2006. We also acknowledge support from the Carleton College Towsley Endowment, as well as the Minnesota Space Grant Consortium.  KC and VS acknowledge partial support by the National Science Foundation through NSF grant AST-2009507

This work has made use of the VALD database, operated at Uppsala University, the Institute of Astronomy RAS in Moscow, and the University of Vienna.

RCT thanks B.~Duffy for his technical assistance with computing resources at Carleton.

We would like to acknowledge that the HET is built on Indigenous land. Moreover, we would like to acknowledge and pay our respects to the Carrizo \& Comecrudo, Coahuiltecan, Caddo, Tonkawa, Comanche, Lipan Apache, Alabama-Coushatta, Kickapoo, Tigua Pueblo, and all the American Indian and Indigenous Peoples and communities who have been or have become a part of these lands and territories in Texas, here on Turtle Island.

\vspace{5mm}
\facilities{10m HET(HPF)}

\software{\texttt{astropy} \citep{2013A&A...558A..33A,2018AJ....156..123A}, 
\texttt{numpy} \citep{5725236},
\texttt{scipy} \citep{2020NatMe..17..261V},
\texttt{HxRGproc} \citep{2018SPIE10709E..2UN},
\texttt{matplotlib} \citep{4160265},
\texttt{GNU Parallel} \citep{tange2011gnu},
\texttt{barycorrpy} \citep{Kanodia2018},
\texttt{SERVAL} \citep{2018A&A...609A..12Z}
          }




\bibliography{KIbibtex.bib}

\end{document}